\documentclass[12pt]{article}

\begin{document}

\title{Space-Time Properties of Hadronization from Nuclear Deep Inelastic Scattering}

\author{W. K. Brooks\thanks{This work is supported in part by the U.S. Department of Energy, including
                DOE Contract No. DE-AC05-84ER40150.}\\*[0.25cm]
        Physics Division\\Thomas Jefferson National Accelerator Facility\\Newport News, Virginia, U.S.A.}
\date{ }

\maketitle

\vspace{-.7cm}

\begin{abstract}
Hadronization, the process by which energetic quarks evolve into
hadrons, has been studied phenomenologically for decades. However,
little experimental insight has been gained into the space-time
features of this fundamentally non-perturbative process. New
experiments at Jefferson Lab, in combination with HERMES data, will
provide significant new insights into the phenomena connected with
hadron formation in deep inelastic scattering, such as quark energy
loss in-medium, gluon emission, and color field restoration.
\end{abstract}

\section{Introduction}

Hadronization has been a ubiquitous feature of high energy physics experiments
for more than two decades. The subculture that has evolved around the
study of this process has succeeded in characterizing the topological
features of high energy jets, such as the fragmentation functions
that describe the probabilities of quarks of a given flavor
evolving into particular hadrons. The string picture is an example of
a successful phenomenology that captures the important
features over a wide range of energies. Few aspects of hadronization,
however, are presently understood at the most fundamental level, since it is a
non-perturbative process.

One important type of experimental data that is generally not easily
accessible for hadronization is its space-time development. While
some phenomenological models can be used to predict this development,
the process takes place on a microscopic distance scale that is not
directly observable. However, this type of information can be inferred by
implanting the hadronization process into a nucleus, which acts as a
spatial filter. Fully formed hadrons in the nuclear medium interact
via ordinary hadronic cross sections; by contrast, propagating quarks
do not appear to interact in a way that modifies hadron 
yields significantly. Therefore, much can be learned from
studying hadron production in deep inelastic scattering kinematics
(DIS) from nuclei of varying sizes. The degree of modification of
fragmentation functions, for instance, can be studied as a function of
nuclear size A, energy transfer $\nu$, hadron energy fraction
$z=E_h/{\nu}$, and the momentum component perpendicular to $\vec{q}$,
which is labelled $p_T$. These studies can yield important insights
into the fundamental processes driving hadronization, such as gluon
emission and the temporal evolution of the truncated color field
following hard interactions.

Several previous measurements of semi-inclusive leptoproduction of
hadrons from nuclei have been performed.
The earliest generation of these were carried out with electron beams
at SLAC\cite{osborne} and muon beams at CERN\cite{ashman} and
FNAL\cite{E665}.
Prominent features displayed by these efforts are an attenuation of
hadrons which increases with A and vanishes with 
increasing $\nu$. Following these early efforts, recent
data\cite{nitrogen}\cite{krypton} from
the HERMES experiment at DESY\cite{hermes}
have inspired a new wave of theoretical interest. Unlike the
historical efforts, these new data offer excellent hadron identification and
high precision, allowing the flavor dependence of hadron formation to
be studied.

Finally, existing and future measurements at Thomas Jefferson National
Accelerator Facility (Jefferson Lab) will offer unique capabilities to
study hadron production in DIS. Measurements at the CEBAF Large
Acceptance Spectrometer (CLAS) at Jefferson Lab\cite{clas}
will provide data on the widest possible range of nuclear target
masses at high luminosity. The first data at 5 GeV were taken in 2003
with sustained luminosities approaching $10^{34} cm^{-2}s^{-1}$,
while 11 GeV data are planned for the future, accompanied by an
order-of-magnitude luminosity increase and improved particle
identification using the upgraded CLAS (``CLAS$^{++}$''). 

The scientific topics connected with these studies are also important
themes in other physics communities. Quark propagation through
strongly interacting media is a foundational element in predicting
signatures of the quark gluon plasma at RHIC through the phenomenon of jet
quenching\cite{wang}; in this connection, it has been explored in
ultrarelativistic d-A collisions\cite{peng}. It has also
been investigated in the Drell-Yan process at FNAL in studies of quark
energy loss in the nuclear medium\cite{garvey}.

\section{Experimental Tools}

A primary experimental tool for nuclear hadron formation is the
\emph{hadronic multiplicity ratio:} 

\vspace{-0.2in}

\begin{eqnarray}
\label{eq:had_mult_rat}
R_M^h(z,\nu,p_T^2,Q^2,\phi)
=
\frac{\Biggl\{ \frac{N_h^{DIS}(z,\nu,p_T^2,Q^2,\phi)}{N_e^{DIS}(\nu,Q^2)}\Biggr\}_A}
{\Biggl\{ \frac{N_h^{DIS}(z,\nu,p_T^2,Q^2,\phi)}{N_e^{DIS}(\nu,Q^2)}\Biggr\}_D}
\end{eqnarray}
where $N_h^{DIS}$ and $N_e^{DIS}$ denote the number of hadrons and
electrons measured in DIS kinematics, $Q^2$ is the four-momentum
transfer, and $\phi$ is the angle between the lepton scattering plane
and the photon-hadron plane. The primary connection to theory has 
traditionally been through $R_M^h$ interpreted in a partonic
framework, for particular flavor combinations. Additional tools
include \emph{energy and angular correlations} between outgoing hadrons
involved in the reaction. There are at least two classes that may be
of interest: correlations between two energetic hadrons formed
directly from the  hadronization of the struck quark, and correlations
between the energetic hadrons and very low energy protons. The
$p_T$ distribution is expected to become broader for larger nuclei; this
\emph{transverse momentum broadening} is expected to be large enough to be
directly measurable\cite{guo}. In addition to
these quantities, \emph{polarization observables} such as single spin
asymmetries may also play a role. 

A variety of physical properties can be derived from the above
observables. Hadron \emph{formation lengths}\footnote{In some models,
referred to as ``production length,'' the distance to produce a
``pre-hadron'' in a two-step picture of hadron formation.}, the characteristic 
distances over which hadrons form, can be extracted from
$R_M^h$. \emph{Quark energy loss} in passing through the nuclear medium may also
be accessible, particularly via transverse momentum
broadening, which is also closely connected with a \emph{quark-gluon
correlation function}. Correlations between final-state particles are
expected to 
shed light 
on the reaction mechanisms, and the nature of the interaction with the
nuclear medium, whether partonic or hadronic.

A variety of experimental
capabilities are required in order to fully exploit the observables
discussed above. Access to the detailed multivariate
dependences of $R_M^h$ requires a large statistical sample; this can
be most economically accessed  
at high luminosity. Isolating the energy and angular correlations
requires large acceptance, since the kinematic variables of the particles of
interest are not strongly correlated. Good particle identification for
several hadron species is required to analyze the flavor dependence of
observables. A large energy range is needed to understand the $\nu$
dependence as well as test the validity of parton model
assumptions. Polarization is needed to study such quantities as single
spin asymmetries. Accessing all of these experimental ingredients
will require combining information from several different facilities.

\section{Model Descriptions and Contact with Observables}

Lattice computations can address nonperturbative processes in
QCD. However, there are several difficulties in applying this
technique to hadronization. A natural choice to describe the time
development of the process is Minkowski space, however, this
introduces some significant technical obstacles. In addition, the
relevant space-time volume needed requires a very large lattice
compared to calculating static properties of hadrons. These
calculations may become feasible in the 5-10 year time frame.

Therefore, to make progress on this topic, modeling is
needed. Most relevant are models possessing close
contact to QCD, and which can predict a wide variety of
observables. A synopsis of model approaches is given in the following
section. 

\subsection{Models}

Historically, a number of models have been developed to describe
hadron production in nuclear DIS. While it is not practical to review
the older models here, a representative approach by Bialas and
Chmaj\cite{bialas} is instructive. In this model, the struck 
quark evolves into a hadron with a characteristic time $\tau$. The
probability that the propagating object is a quark at a given time $t$
after it is struck is $P_q=e^{-t/\tau}$ and the corresponding
probability that it has become a hadron is simply
$P_h=1-P_q$. Phenomenological cross sections for the quark and hadron
interacting with the nuclear medium are postulated: $\sigma_q$ and
$\sigma_h$. The probability of the quark or hadron interacting with the
nuclear medium is taken as proportional to the density of the medium
for a spherical nucleus, using a standard density
parameterization. $R_M^h$ can then be calculated by numerical
integration. This model approach was qualitatively successful in 
describing the $\nu$ dependence of $R_M^h$ for the HERMES nitrogen
data\cite{nitrogen}, but failed to describe the $z$ dependence even
qualitatively.

In a more sophisticated approach, the gluon bremsstrahlung 
model\cite{boris} divides the hadronization process into
several 
stages. As the struck quark propagates, an initial stage of
uninhibited gluon emission occurs, followed by a second 
stage of reduced gluon emission. The suppression of gluon emission
in the second stage is imposed to conserve energy, since the
propagating quark begins with only a finite amount of energy,
$\nu$. At some point 
in the second stage, a color dipole is formed from the struck quark
and the last emitted gluon. The color dipole subsequently evolves into
a hadron. The role of the nuclear medium is to modulate the rate at
which the color dipole turns into a hadron, including the effect of
color transparency; in later versions of the model, medium-stimulated
gluon emission is included, which has a small effect on the prediction
for $R_M^h$. The assumptions of the model are valid for pions with 
$z>0.5$. Its predictions agree well with the HERMES nitrogen data in
both the $z$ dependence and the $\nu$ dependence.

A more recent model approach to describing hadron production in
nuclear DIS is based on a continuous series of developments since the late 1980's
addressing jet quenching in relativistic heavy ion 
reactions\cite{wang}. In this approach, the prediction
for the 
behavior of $R_M^h$ is based solely on medium-stimulated gluon
emission; hadronization is assumed to occur outside the nuclear
medium. The calculation of $R_M^h$ is performed in pQCD; in addition
to the leading order terms, the dominant twist-four term is also
included. The twist-four term represents multiple scattering in the
nuclear medium, and the contribution due to this term is a free parameter
that can be fixed by comparison to data. This model was able to
describe the HERMES krypton data by fitting the HERMES
nitrogen data. The modification of the fragmentation functions
represented by $R_M^h(z,\nu)$ emerges naturally from this
description. In this approach, a novel quadratic dependence of the
energy loss on path length is predicted\cite{bdmps}, originating in the
non-Abelian analog of the LPM effect in QED\cite{lpm}. However, the
assumption that hadronization occurs outside the nucleus will break
down at low energies or for more massive hadrons. 

Another recent model invokes deconfinement of nucleons in the
nuclear medium as a mechanism for the modification of
fragmentation functions in the nuclear medium\cite{pirner}. The basic method follows the rescaling models developed in
the 1980's. This approach combines partial deconfinement with
absorption of the formed 
hadron in the nuclear medium. The nucleonic deconfinement effectively
extends the range in $Q^2$ for which gluon radiation takes place,
modifying the fragmentation functions accordingly. The model
successfully describes the EMC data and the 
HERMES data for pions and $K^+$ in nitrogen and krypton, although the
prediction for $K^-$ systematically deviates from the data.

A fourth recent model adapts and improves stationary string model
(SSM) techniques developed in the 1980's and 1990's. This model,
applied to valence quarks, successfully
fits the HERMES pion data for nitrogen and krypton, and predictions
are given for kaons and heavier 
nuclei\cite{armenian}. As with the previous model, the
prediction for $K^+$ and $K^-$ attenuations are very similar, while the
HERMES data show a systematic difference. 

Finally, an approach that emphasizes accurate modeling of the final
state interactions (FSI) is able to explain the HERMES data without invoking
modification of fragmentation functions\cite{mosel}. The
struck quark propagates, evolves into a hadron, and its interaction with the
nuclear medium is given by a semi-classical coupled channel transport
calculation that has been tested in other reactions. These
calculations, although in preliminary form, seem to have enough
flexibility to reproduce the HERMES nitrogen and krypton
data for charged pions and kaons, and protons and anti-protons. There
is good qualitative agreement throughout, and quantitative agreement for
some kinematics. 

\subsection{Connections between Models and Observables}

The models highlighted above are generally able to describe the
published data, in spite of the fact that they rely on quite
different physical pictures. In order to make good progress in
understanding hadron formation in nuclear DIS, it is necessary to
discriminate among these models using more data that span a wider
range of kinematics or that introduce new observables. Similarly, more
complete calculations are needed. For instance, model predictions for
the flavor and mass dependence of $R_M^h(z,\nu)$ can be tested by
studying pions, kaons, and protons. In fact, a wide range of particles can
be considered, as seen in Table \ref{table:hadron_list}. In this table,
hadrons with 
$c\tau$ much greater than nuclear dimensions are listed. To the extent
that hadronization is the production mechanism, inter-comparison of
$R_M^h(z,\nu)$ for these particles will shed light on
such topics as whether semi-classical calculations of FSI suffice, or
whether a quantum mechanical description is 
required. Likewise, study of moments of transverse momentum may
distinguish between models with different assumptions about the number
of multiple scatters or the shape of the emitted gluon momentum
distribution. Measurements of $R_M^h$ at low $z$ may be useful,
particularly at low energies, to distinguish between pion flux
entering other reaction channels, as in FSI, from a change in the
intrinsic fragmentation function, i.e., due to gluon emission. A clean
evaluation of the $Q^2$ dependence of $R_M^h$ will 
help distinguish between different reaction mechanisms. Studies of the
angular correlations between two high-energy pions may also clarify
whether the reaction mechanism is dominated by FSI in the nuclear
medium or by the fragmentation process. Spin observables, such as
single-spin asymmetries, may offer indications of whether partonic
multiple scattering is an important component of the process.
Finally, the twist-4 pQCD model makes the assumption that hadronization takes
place outside the nucleus altogether. The point at which this 
assumption is no longer valid is an important unknown that can be
addressed by the CLAS low-energy data.

\begin{table}[h!tb]
\begin{caption}
	{\label{table:hadron_list}\small Final-state hadrons potentially
	accessible for formation 
	length and transverse momentum broadening
	studies in CLAS. The rate estimates were obtained from the LEPTO
	event generator for an 11 GeV incident electron beam. (The
	criteria for selection of these particles was that $c\tau$ should
	be significantly larger than nuclear dimensions, and their decay
	channels should be measurable by CLAS$^{++}$.)}
\vspace{3mm}
\end{caption}
\begin{center}
\begin{tabular}{|c|c|c|c|c|c|}

\hline
\hline
hadron & $c\tau$ & mass & flavor  & detection & production rate\\
       &         &(GeV) & content &  channel &  per 1k DIS events\\
\hline
\hline
$\pi^0$ & 25 nm & 0.13 & $u\bar{u}d\bar{d}$ & $\gamma\gamma$ & 1100 \\
\hline
$\pi^+$ & 7.8 m & 0.14 &   $u\bar{d}$ & direct & 1000 \\
\hline
$\pi^-$ & 7.8 m & 0.14 &   $d\bar{u}$  & direct & 1000 \\
\hline
$\eta$ & 0.17 nm & 0.55 & $u\bar{u}d\bar{d}s\bar{s}$&$\gamma\gamma$ &
120 \\
\hline
$\omega$ & 23 fm & 0.78 &  $u\bar{u}d\bar{d}s\bar{s}$ &
$\pi^+\pi^-\pi^0$ & 170\\
\hline
$\eta'$ & 0.98 pm & 0.96 &  $u\bar{u}d\bar{d}s\bar{s}$ &
$\pi^+\pi^-\eta$ & 27 \\
\hline
$\phi$ & 44 fm & 1.0 &  $u\bar{u}d\bar{d}s\bar{s}$ & $K^+K^-$ & 0.8 \\
\hline
$K^+$ & 3.7 m & 0.49 &  $u\bar{s}$ & direct & 75\\
\hline
$K^-$ & 3.7 m & 0.49 &  $\bar{u}s$ & direct & 25\\
\hline
$K^0$ & 27 mm & 0.50 &  $d\bar{s}$ & $\pi^+\pi^-$ & 42 \\
\hline
$p$ & stable & 0.94 &  $ud$ & direct & 1100 \\
\hline
$\bar{p}$ & stable & 0.94 &  $\bar{u}\bar{d}$ & direct & 3 \\
\hline
$\Lambda$ & 79 mm & 1.1 &  $uds$ & $p\pi^-$ & 72 \\
\hline
$\Lambda(1520)$ & 13 fm & 1.5 &  $uds$ & $p\pi^-$ & - \\
\hline
$\Sigma^+$ & 24 mm & 1.2 &  $us$ & $p\pi^0$ & 6 \\
\hline
$\Sigma^0$ & 22 pm & 1.2 &  $uds$ & $\Lambda\gamma$ & 11 \\
\hline
$\Xi^0$ & 87 mm & 1.3 &  $us$ & $\Lambda\pi^0$ & 0.6 \\
\hline
$\Xi^-$ & 49 mm & 1.3 &  $ds$ & $\Lambda\pi^-$ & 0.9 \\
\hline
\hline	
\end{tabular}
\end{center}
\end{table}

While these are rich and exciting prospects for learning much new
physics, there are a few experimental and theoretical complications
that require 
further study and clarification. For instance, hadron formation
lengths can be derived from the data, however, the definition of
the formation length depends on which model is considered. As an 
example, hadron formation is conceived as a multi-step process with multiple
time constants in some models, but a single-step process in
others. Reference frame considerations also enter the interpretation
of the data; a lower limit on x is required to preserve the simple
picture that a single quark propagates with initial energy
$\nu$\cite{brodsky}. Resonances in the residual system may need to
be suppressed, which reduces the kinematic coverage available.
There may be an issue of isolating current fragmentation from
target fragmentation, although the nuclear medium may perhaps suppress
the latter. Finally, the degree of validity of QCD factorization needs
to be evaluated. The last few issues are common to all
semi-inclusive deep inelastic scattering, and as such are being
investigated by several physics communities; further progress is
therefore to be expected. 

\section{Future Prospects}
Over the next few years it is expected that new datasets will become
available which will provide further stringent tests of these
models. Some HERMES data were taken with a 12 GeV beam on both 
nitrogen and krypton targets; 27 GeV data on helium and neon are also
under 
analysis. A new program at Jefferson Lab will provide even lower energy
data at high luminosity on targets ranging from carbon to lead. This
5 GeV data, from the CLAS EG2 running period, will have sufficient
statistics to map out the full multivariate character of $R_M^h$ for
several hadron species, and to directly measure transverse momentum
broadening. 

Other future prospects include the possibility of additional targets
in HERMES such as xenon, to complete the study of nuclear
mass dependence at high energy transfer, and a large program which
would be possible at Jefferson Lab following the energy upgrade. The
ultimate moderate-energy program will be feasible with CLAS$^{++}$, due to
the planned upgrade in luminosity (to $10^{35} cm^{-2}s^{-1}$) and
improved particle identification capabilities. These improvements will
permit mapping out multivariate $R_M^h$ for all the hadrons in Table 1
as well as $p_T$ broadening for many of these, spanning a kinematic
range of $\nu$ from $2-9 ~GeV$ and $2-9 ~GeV^2$ in $Q^2$.  

\section{Conclusions}
The study of hadronization from semi-inclusive nuclear DIS is a new
topic that is drawing 
increasing interest from several different physics communities.
The recent wave of interest stimulated by the HERMES measurements has
resulted in 
a number of theoretical efforts to describe the data. These efforts,
based on a variety of physical pictures, demonstrate the exciting
possibility of gaining new insight into hadronization as well as other
aspects of quark propagation through nuclei such as quark energy
loss. However, more data is needed
to definitively discriminate among these model approaches. In
addition, much more theoretical effort is needed to address all of the
recent data, to reduce the number of assumptions in the models, and to
make greater contact with the fundamental theory, QCD.

\newpage 

\end{document}